\documentclass[aps, twocolumn, prl, citeautoscript, superscriptaddress, longbibliography, a4paper]{revtex4-2}

\usepackage{amssymb,amsfonts,amsmath,amsthm}
\usepackage{graphicx}
\usepackage[english]{babel}
\usepackage[latin1]{inputenc}
\usepackage[pdftex,colorlinks=true,pdfstartview=FitH,linkcolor=blue,citecolor=blue,urlcolor=blue]{hyperref}

\usepackage{bm}
\usepackage{float}
\usepackage{mathtools}
\usepackage{color,soul}
\usepackage{upgreek}
\usepackage{MnSymbol}
\usepackage{verbatim}
\usepackage[bottom]{footmisc}
\usepackage{bbold}

\usepackage[dvipsnames,svgnames,table]{xcolor}
\usepackage{hyperref}
\usepackage{fancyhdr}
\usepackage[english]{babel}
\usepackage[caption=false]{subfig}

\hypersetup{
pdfstartview={FitH},
colorlinks=true,    
linkcolor=NavyBlue, 
citecolor=Blue,   
filecolor=NavyBlue, 
urlcolor=NavyBlue   
}

\newcommand{\bea}{\begin{eqnarray}}
\newcommand{\eea}{\end{eqnarray}}
\newcommand{\be}{\begin{equation}}
\newcommand{\ee}{\end{equation}}

\newcommand{\dt}[1]{\frac{d #1}{dt} }

\begin{document}
\title{Shapiro resonances in ac-self-modulated exciton-polariton Josephson junctions}

\author{I. Carraro Haddad}
\affiliation{Centro At{\'{o}}mico Bariloche and Instituto Balseiro,
Comisi\'on Nacional de Energ\'{\i}a At\'omica (CNEA)- Universidad Nacional de Cuyo (UNCUYO), 8400 Bariloche, Argentina.}
\affiliation{Instituto de Nanociencia y Nanotecnolog\'{i}a (INN-Bariloche), Consejo Nacional de Investigaciones Cient\'{\i}ficas y T\'ecnicas (CONICET), Argentina.}

\author{D. L. Chafatinos}
\affiliation{Centro At{\'{o}}mico Bariloche and Instituto Balseiro,
Comisi\'on Nacional de Energ\'{\i}a At\'omica (CNEA)- Universidad Nacional de Cuyo (UNCUYO), 8400 Bariloche, Argentina.}
\affiliation{Instituto de Nanociencia y Nanotecnolog\'{i}a (INN-Bariloche), Consejo Nacional de Investigaciones Cient\'{\i}ficas y T\'ecnicas (CONICET), Argentina.}


\author{A.~A. Reynoso}
\affiliation{Centro At{\'{o}}mico Bariloche and Instituto Balseiro,
Comisi\'on Nacional de Energ\'{\i}a At\'omica (CNEA)- Universidad Nacional de Cuyo (UNCUYO), 8400 Bariloche, Argentina.}
\affiliation{Instituto de Nanociencia y Nanotecnolog\'{i}a (INN-Bariloche), Consejo Nacional de Investigaciones Cient\'{\i}ficas y T\'ecnicas (CONICET), Argentina.}

\author{A. E. Bruchhausen}
\affiliation{Centro At{\'{o}}mico Bariloche and Instituto Balseiro,
Comisi\'on Nacional de Energ\'{\i}a At\'omica (CNEA)- Universidad Nacional de Cuyo (UNCUYO), 8400 Bariloche, Argentina.}
\affiliation{Instituto de Nanociencia y Nanotecnolog\'{i}a (INN-Bariloche), Consejo Nacional de Investigaciones Cient\'{\i}ficas y T\'ecnicas (CONICET), Argentina.}

\author{A.~S. Kuznetsov}
\affiliation{Paul-Drude-Institut f\"{u}r Festk\"{o}rperelektronik, Leibniz-Institut im Forschungsverbund Berlin e.V., Hausvogteiplatz 5-7,\\ 10117 Berlin, Germany.}

\author{K. Biermann}
\affiliation{Paul-Drude-Institut f\"{u}r Festk\"{o}rperelektronik, Leibniz-Institut im Forschungsverbund Berlin e.V., Hausvogteiplatz 5-7,\\ 10117 Berlin, Germany.}

\author{P.~V. Santos}
\affiliation{Paul-Drude-Institut f\"{u}r Festk\"{o}rperelektronik, Leibniz-Institut im Forschungsverbund Berlin e.V., Hausvogteiplatz 5-7,\\ 10117 Berlin, Germany.}

\author{G. Usaj}
\affiliation{Centro At{\'{o}}mico Bariloche and Instituto Balseiro,
Comisi\'on Nacional de Energ\'{\i}a At\'omica (CNEA)- Universidad Nacional de Cuyo (UNCUYO), 8400 Bariloche, Argentina.}
\affiliation{Instituto de Nanociencia y Nanotecnolog\'{i}a (INN-Bariloche), Consejo Nacional de Investigaciones Cient\'{\i}ficas y T\'ecnicas (CONICET), Argentina.}

\author{A. Fainstein}
\email[Corresponding author, e-mail: ]{alejandro.fainstein@ib.edu.ar}
\affiliation{Centro At{\'{o}}mico Bariloche and Instituto Balseiro,
Comisi\'on Nacional de Energ\'{\i}a At\'omica (CNEA)- Universidad Nacional de Cuyo (UNCUYO), 8400 Bariloche, Argentina.}
\affiliation{Instituto de Nanociencia y Nanotecnolog\'{i}a (INN-Bariloche), Consejo Nacional de Investigaciones Cient\'{\i}ficas y T\'ecnicas (CONICET), Argentina.}

\date{\today}

\begin{abstract}
{
We experimentally investigate the dynamics of exciton polariton Josephson junctions when the coupling between condensates is periodically modulated through self-induced mechanical oscillations. The condensates energy detuning, the analog of the bias voltage in superconducting junctions, displays a plateau behavior akin to the Shapiro steps. At each step massive tunneling of particles occurs featuring Shapiro-like spikes. These characteristic changes are observed when the condensates Josephson frequency $\omega_\textrm{J}$ is an integer multiple of the modulation frequency $\omega_\mathrm{M}$. 
}
\end{abstract}
\maketitle


\textit{Introduction.---}Macroscopic tunneling through a barrier is a paradigm of quantum mechanics originally studied by Josephson in the context of superconductor-insulator-superconductor junctions (SJJ) \cite{Josephson1962}. As established in his original proposal, a time independent quantum phase difference $\phi$ across the junction  establishes a dissipationless electrical dc current $I$, given by the relation $I = I_{c} \sin\phi$, with $|I|\leq I_c$ the critical current (the dc Josephson effect).  The time dependence of $\phi$ is determined by the chemical potential $\Delta \mu$ applied across the junction, $\dot{\phi} = - \frac{\Delta \mu}{\hbar}$. In a superconductor, $\Delta \mu = -2eV$ ($2e$ is the Cooper pair charge) and, thus, the application of a dc voltage $V$ produces current oscillations at the Josephson frequency $\omega_\textrm{J} = 2eV/\hbar$ (the ac Josephson effect). Besides superconducting junctions, the dc and ac Josephson effects have been observed in other systems characterized by macroscopic phase coherent wave functions such as two superfluid helium reservoirs connected by nanoscopic apertures~\cite{Backhaus1998}, coupled Bose-Einstein cold-atom condensates defined by engineered electromagnetic potentials~\cite{Smerzi1997,Giovanazzi2000,Levy2007}, or coupled exciton-polariton condensates induced in microstructured or light-induced traps~\cite{Lagoudakis2010,CarusottoRMP2013,Voronova2025}. These systems are usually also characterized by the relevance of non-linearities (particle-particle interactions) and dissipation, which has shown to lead to peculiar phenomena. For example, interactions at high densities can quench the transfer of particles across the junction inducing a macroscopic self-trapping  on one side as observed for both atomic and polariton condensate junctions~\cite{Albiez2005,Abbarchi2013}.

When a superconducting Josephson junction is exposed to microwave radiation of frequency $\omega$, quantized voltage steps occur when $\omega_\textrm{J} = n \omega$ (Shapiro steps)~\cite{Shapiro1963}. A related effect was later also observed for superfluid $^3$He weak links under the action of an oscillating pressure~\cite{Simmonds2001}. Very recently, experiments and theoretical proposals have been presented for its observation in coupled atomic condensates with a moving vibrating potential boundary~\cite{Grond2011,Singh2024,DelPace2024,Bernhart2024}.

In this work we report the observation of the analog of the Shapiro effect (also identified as inverse ac Josephson effect) in coupled exciton-polariton condensates under a periodic mechanical GHz drive. In contrast to all previous implementations, the periodic perturbation is self-induced through an optomechanical process launched by the two Josephson-coupled polariton condensates which are slightly detuned. Depending on the physical parameters, single or multiple attractors can coexist in the dynamical phase diagram of the system. 

\textit{Polariton Josephson junctions with phonon confinement.---}
Our optomechanical platform consist of two confined exciton-polariton Bose-Einstein condensates (BECs)~\cite{CarusottoRMP2013}, weakly linked through tunneling (coupling $J$).
Microcavity exciton-polaritons (hereafter referred to as polaritons) are composite quasi-particles resulting from the strong coupling between excitons and confined cavity photons~\cite{CarusottoRMP2013}. Owing to their photonic component, polaritons exhibit a low effective mass and inherit the propagation characteristics of photons within the cavity.  
Due to their exciton component, polaritons couple through Coulomb interactions between themselves (interaction $U_0$) and also with the exciton reservoir ($U_\textrm{R}$). 
Polaritons can decay to photon states outside the microcavity with a decay rate $\gamma$, and be thus detected using photoluminescence (PL) spectroscopy.
Due to stimulated scattering and above a laser pumping threshold, exciton-polaritons form a coherent macroscopic quantum state reminiscent of BECs~\cite{Kasprzak2006,Balili2007,Amo2009}.

Distributed Bragg reflector microcavities based on GaAs/AlAs can be designed to confine both infrared light and acoustic vibrations in the GHz range~\cite{Trigo2002,Fainstein2013}. For optical cavities tuned to resonance with GaAs quantum well excitons, the fundamental frequency of the confined acoustic vibrations falls around  $\nu^0_\textrm{M} = \omega^{0}_\textrm{M}/2\pi \sim 20$~GHz, and higher overtones occur at $\nu^{(n)}_\textrm{M} = (1+2n) \nu^{0}_\textrm{M} \sim 60, 100,...$~GHz ($n=1,2...$). Due to their exciton component, polaritons also exhibit strong interactions with phonons, leading to optomechanical coupling factors $g_0/2\pi \sim 20$~MHz~\cite{Rozas2014,Jusserand2015,Zambon2022,Sesin2023,Santos2023} which are much larger than the record values in photon optomechanics~\cite{RMP}. The polariton Josephson junctions (PJJs) studied here were fabricated by micro-structuring two square traps of lateral size  $4\,\mu$m, separated by either $1\,\mu$m or $2\,\mu$m. 
The micro-structuring process involves etching the spacer of the  microcavities around the traps, locally altering the spacer thickness and consequently modifying the energy levels of cavity photons and phonons~\cite{Winkler2015,Kuznetsov2018,Chafatinos2022}.  When these structures are designed to display polariton levels with energy splittings matching the phonon frequencies, efficient self-induced mechanical oscillations (phonon lasing)~\cite{Kippenberg2005,Grudinin2010} is observed~\cite{Chafatinos2020,Reynoso2022}. As we show here, this modulation resembles the application of an ac field in SJJs, and leads to the resonant tunneling of polaritons between condensates, resulting in phase-locked steps akin to the Shapiro effect.

\begin{figure}[ht!]
    \centering 
\includegraphics[trim = 0mm 0mm 0mm 0mm, clip=true, keepaspectratio=true, width=0.75\columnwidth,angle=0]{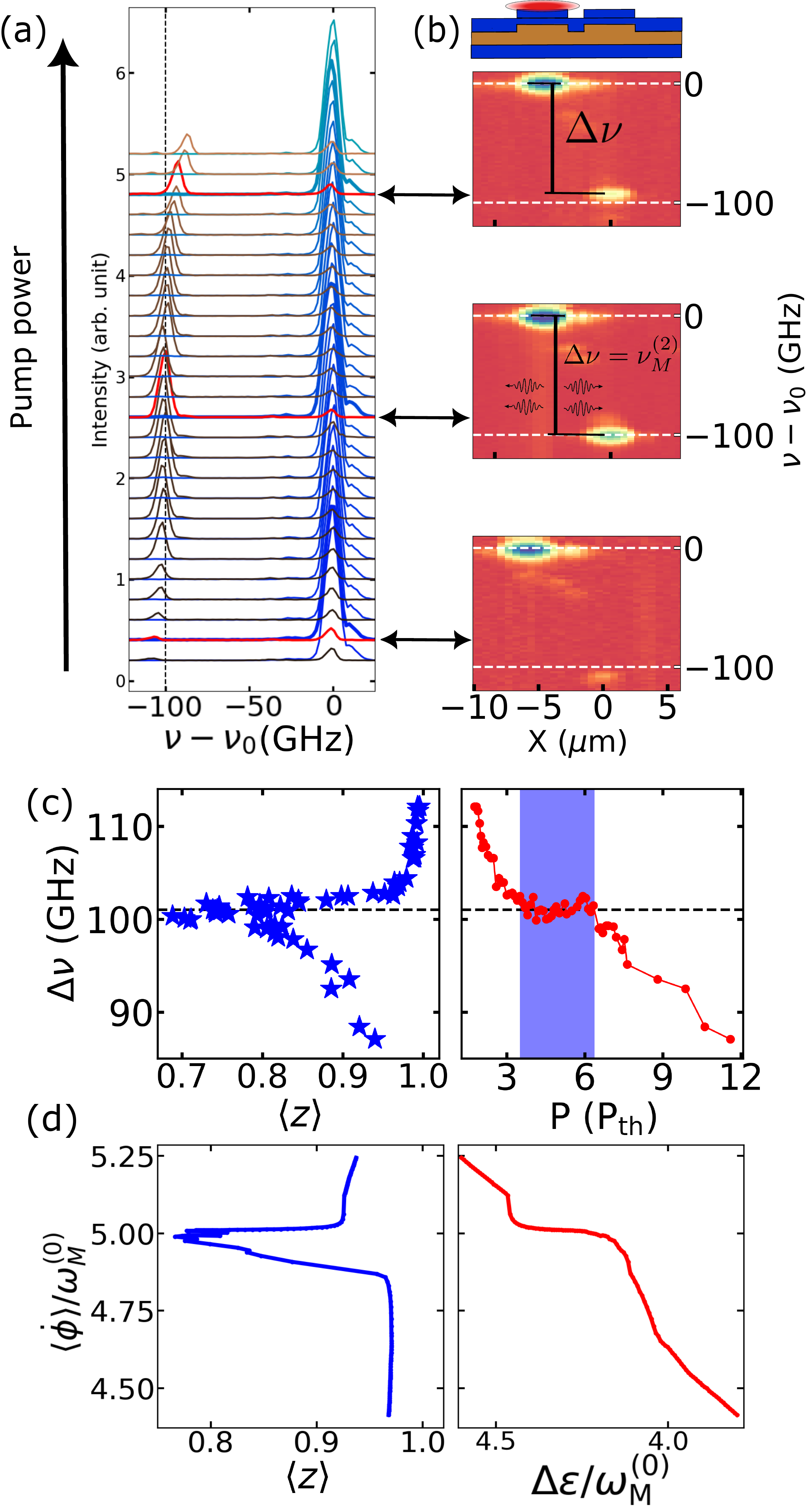}
 \caption{Shapiro steps and spikes of a mechanically modulated PJJ.
(a) Cascade of PL spectra for increasing non-resonant excitation power (from bottom to top) for a PJJ made of two square traps of $4\mu$m lateral size separated by $2\mu$m. (b) Spectrally resolved spatial images corresponding to three selected spectra highlighted in (a). Note that $\sim 100$~GHz is the frequency of a confined vibration in the structure. (c) Measured frequency separation between the trap levels $\Delta \nu$ as a function of applied power $P$, and the corresponding population imbalance $\langle z \rangle$ derived from the spectra in (a). (d) Modeling of (c) based on the full gGPEs for the PJJ (see text for details). Here the time averaged $\langle \dot\phi \rangle$ is a measure of the dressed detuning while $\Delta{\varepsilon}$ is the bare detuning.
}
\label{Fig2}
\end{figure} 

\begin{figure}[t]
 \begin{center}
    \includegraphics[trim = 0mm 0mm 0mm 0mm,clip=true, keepaspectratio=true, width=1 \columnwidth,angle=0]{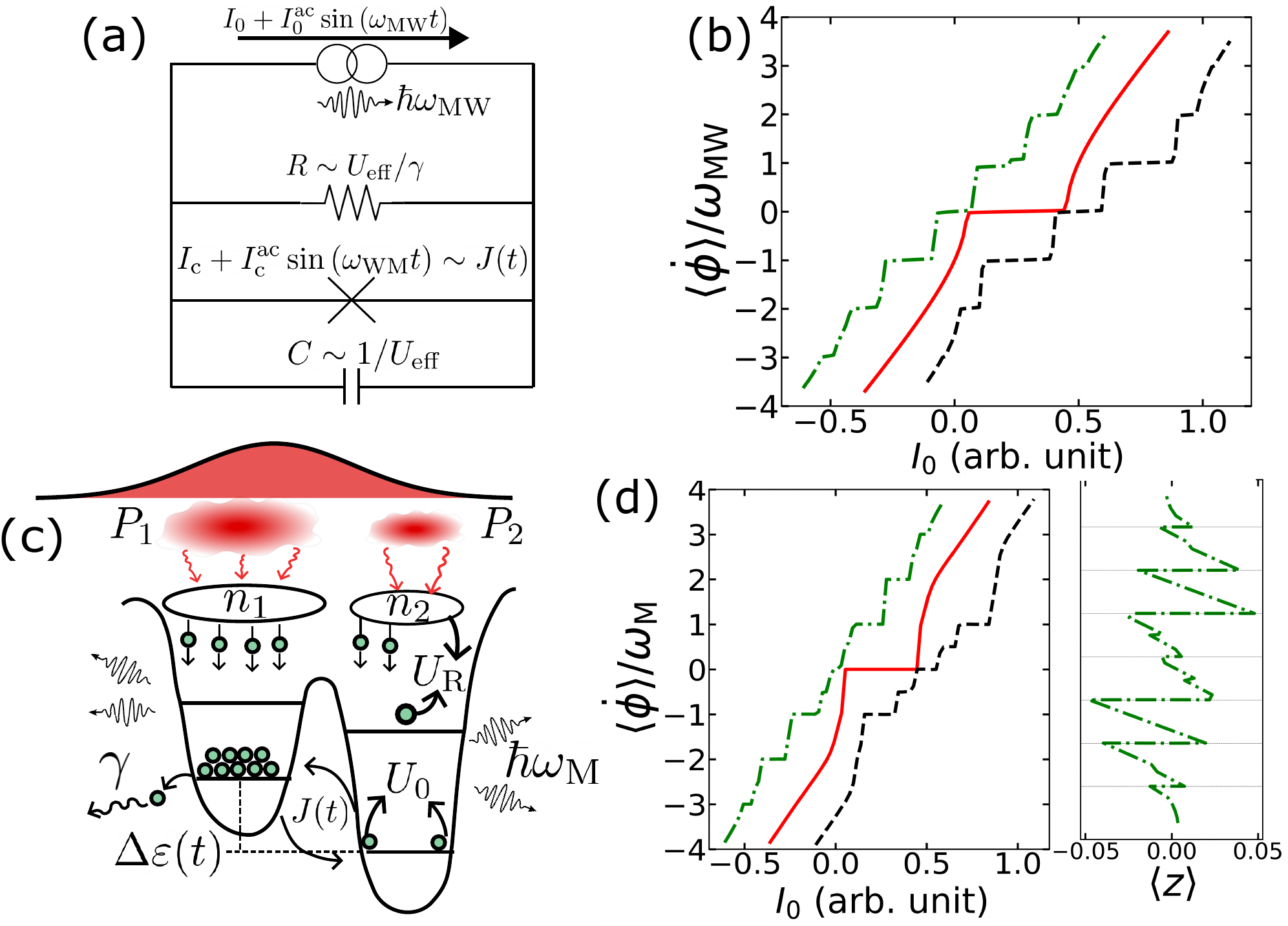}
\end{center}
\vspace{-0.5 cm}
\caption{(a) Scheme of the equivalent RCSJ circuit. (b) Phase derivative $\langle \dot\phi \rangle$ as a function of the ac driving current $I_0$ for the RCSJ model: current dc driving $I_0^\mathrm{ac}=I_{c}^\mathrm{ac}=0$ (solid red curve), current ac driving $I_0^\mathrm{ac}\neq0$ (left green dashed-dotted curve) and ac modulation of the critical current $I_c^\mathrm{ac}\neq0$ (right black dashed). These curves are horizontally shifted for clarity. (c) Scheme of the double trap PJJ (see text). (d) Left panel, same as (b) but for the case of a PJJ modulated by mechanical oscillations of frequency $\omega_\textrm{M}$ (data obtained using the full gGPE). Right panel, population imbalance $\langle z \rangle$ as a function of the dressed detuning $\langle \dot\phi \rangle$ with ac driving of the energy detuning $\Delta \varepsilon(t)$.
}
\label{Fig1}
\end{figure}

 
\textit{Experimental evidence of Shapiro steps and spikes in PJJs.---}Figure~\ref{Fig2} shows the PL spectrum of two coupled $4 \times 4 \mu$m$^2$ traps separated by 2 $\mu$m as a function of the excitation power (increasing from bottom to top). The energy difference between the levels of the traps is varied by placing the non-resonant excitation (with a $760$~nm laser and a $\sim 5 \mu$m diameter spot) displaced towards one of the traps. The population imbalance between the traps alters their relative energy through the energy blue shift induced by the interaction between polaritons and between polaritons and reservoir excitons. This blue-shift saturates above a certain power, so that the less-pumped trap starts to catch-up decreasing the energy detuning~\cite{Chafatinos2020}. Panel (a) displays the power dependence of spectra corresponding to the  polariton ground state (GS) of the traps. Blue (brown) color is used for the light spatially collected from the left (right) more (less) pumped trap.  Energies are given relative to the most intense peak, to discount the  overall blue shift of the spectra. The relative GS energy of the less pumped trap continuously increases but, notably, for a range of applied powers, it locks to a detuning close to $100$~GHz. We recall that this latter value coincides with the second overtone ($n=2$) of the confined phonon frequency~\cite{Fainstein2013}. Moreover, as shown with the spatially and spectrally resolved emission on the color maps in Fig.~\ref{Fig2}(b), the relative intensity of this shifting mode clearly increases when the locking occurs. 
The measured frequency separation between trap modes $\Delta \nu$ as a function of the population imbalance  $\langle z \rangle$ and the laser power are presented in Fig.~\ref{Fig2}(c). Clearly, when the detuning approaches $\sim 100$~GHz, the population imbalance decreases, indicating a massive transfer of population to the non-pumped trap. Concomitantly, $\Delta \nu$ locks to $\sim 100$~GHz over a large range spanning approximately from $3.5$ to $6\,P_\textrm{th}$ (highlighted by the blue area in the figure). Here $P_\textrm{th}$ denotes the condensation threshold pump power. These experimental features resemble the physics of Shapiro steps and spikes observed in superconducting Josephson junctions.


\textit{The RCSJ model.---}Interactions and dissipation are present in superconducting junctions when the applied voltage and temperature are finite. The physics of the superconducting Josephson junction in this case is captured by the resistively and capacitively shunted junction (RCSJ) model~\cite{Stewart1968,McCumber1968}. This model considers the effect of leakage currents through the insulator barrier, which act as an effective resistance $R$, as well as a capacitive-like effect ($C$) through the buildup of charge at the interface layers of the junction---the effective parallel circuit is depicted in Fig.~\ref{Fig1}(a). This circuit is described by the equations
\begin{equation}
I = \frac{\hbar C}{2e} \ddot{\phi} + \frac{\hbar}{2eR} \dot{\phi} + I_c \sin{\phi}\,,\qquad
 \dot{\phi} = \frac{2e}{\hbar}V\,,
\label{RCSJ}
\end{equation}
where the total current $I$ is the sum of the three contributions: capacitive, resistive/dissipative, and the coherent tunneling (Josephson).
%
Typical solutions of Eq.~\eqref{RCSJ} are illustrated in Fig. \ref{Fig1}(b). When current biased with a dc current $I_0$, 
a dissipation-less (flat) region exists where there is no voltage. If the current surpasses the critical value $I_c$, the Josephson junction transitions to a state in which a measurable dc voltage drop appears. This $I$-$V$ curve can be understood as a classical bifurcation from a synchronized state (flat region) to Josephson dynamics~\cite{Zibold2010}. When the junction is irradiated with microwaves of frequency $\omega_\textrm{MW}$, an alternating current is induced resulting in $I(t) = I_0 + I_0^\textrm{ac}\cos{\omega_\textrm{MW}t}$. Under these conditions, if the dc component of the phase derivative approaches a harmonic of the driving frequency, $\langle \dot{\phi} \rangle=n\omega_\textrm{MW}$, it becomes locked to that particular frequency. This locking phenomenon, the so-called `Shapiro steps'~\cite{Shapiro1963}, manifests as a staircase pattern when $\langle \dot{\phi} \rangle$ is plotted against $I_0$. Figure~\ref{Fig1}(b) illustrates the response of $\langle \dot{\phi} \rangle$ when either the external current $I$ (dot-dashed green curve) or the critical current $I_\textrm{c}$ (black-dashed curve) are harmonically modulated in time.

\begin{figure}[ht!]
    \centering 
    \includegraphics[trim = 0mm 0mm 0mm 0mm,clip=true, keepaspectratio=false, width=0.9\columnwidth,angle=0]{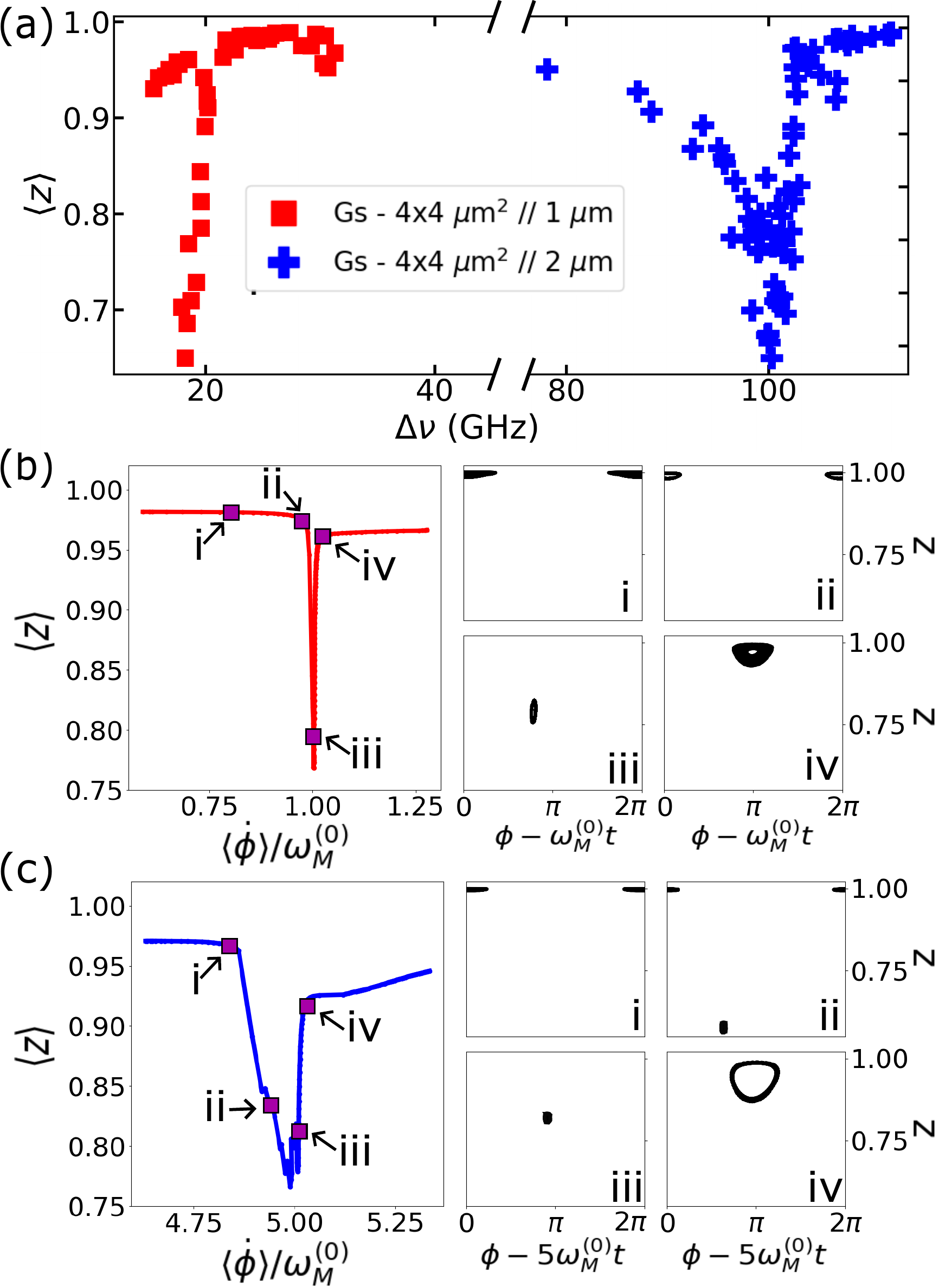}
    \caption{Optomechanically-induced Shapiro resonances in PJJs. 
    (a) Shapiro spikes (population imbalance $\langle  z \rangle$ vs inter-level frequency detuning $\Delta \nu$, for two PJJs of different trap separation and mechanical resonance frequency. Note the different shape of the spikes.
    (b-c) Modeling of Shapiro spikes with the full gGPEs for two situations displaying bifurcations involving single (b) or double (c) attractors. These attractors are shown for both (b) and (c) at four values of $\langle \dot\phi \rangle $, identified with purple squares. The change from single to double attractor phases correspond to the points identified as $(ii)$, single in (b) and double in (c). Calculations in panel (b)[(c)] are obtained with $\gamma = 0.42$ ($0.15$), $\nu_\textrm{M} = 1$ ($5$), $J_0 = 0.05$, $J_\mathrm{M} = 0.07(0.05)$, $U_0 = 0.7$, $U_\mathrm{R} = 2.6$, $\gamma_R = 0.45$, $p_1 = 2.8$, $p_2 = 1.0$. Parameters are referred to the fundamental confined phonon frequency $\nu^0_\mathrm{M} = 20$~GHz, and the $p_i$ to the condensation threshold $P_\mathrm{th}$.
 }
    \label{Fig3}
\end{figure}
 
\textit{Equivalence between the RCSJ and generalized Gross Pitaevskii models of a junction.---}
The modeling of our PJJ is  schematized in Fig~\ref{Fig1}(c). The selective pumping of two optical traps with power $p_i$ ($i=1,2$)  creates exciton reservoirs of population $n_i$ that in turn feed the polariton traps. The resulting coupled polariton condensates can be described by  generalized Gross-Pitaevskii equations (gGPEs) \cite{Wouters2007} for the complex functions ${\psi}_j(t)$ ($j = 1,2$) that represent the amplitude of each trapped mode of bare energy $\varepsilon_j$ (see \cite{SM}). 
Writing ${\psi}_j (t)= \sqrt{N_j(t)} e^{i\theta_j(t)}$ and introducing the relative variables  $\phi(t) = \theta_1(t) - \theta_2(t)$, the phase difference between condensates, and $z(t) = (N_1(t) - N_2(t))/(N_1(t) + N_2(t))$, the population imbalance, it is straightforward~\cite{SM} to show that the gGPEs reduce to
\begin{eqnarray}
\label{zphi}
\dot{z} &=& - \gamma z + \frac{2J}{\hbar} \sqrt{1-z^2}\sin{(\phi)}\,,\\
\nonumber
\dot{\phi} &=& -\frac{\Delta\varepsilon}{\hbar} - 2\frac{U_0}{\hbar} p z + 2 \frac{U_R}{\hbar} \frac{z}{1-z^2}- \frac{2J}{\hbar} \frac{z}{\sqrt{1-z^2}} \cos{(\phi)}\,,\\
\nonumber
\end{eqnarray}
with $\Delta \varepsilon = \varepsilon_1 - \varepsilon_2$ and where, without loss of generality, and to simplify the expressions, we assumed a balanced pumping of the two reservoirs, i.e. $p_1 = p_2 = p$. 
%
Finally, for weakly coupled condensates, $|J|\ll |U_0 p-U_R|$, one can consider the current between them to be relatively small and so the population imbalance $|z| \ll 1$. In this limit, the following simplified model is derived~\cite{SM}
\begin{eqnarray}
\nonumber
-\frac{\gamma \Delta \varepsilon}{2 U_\mathrm{eff}}  &=& \frac{\hbar}{2U_\mathrm{eff}} \ddot{\phi} + \frac{\hbar \gamma}{2U_\mathrm{eff}} \dot{\phi}    +s \frac{2J}{\hbar} \sin{\phi}\,,\\
\dot{\phi} &=& 
-\frac{2(U_0p-U_R)}{\hbar} z - \frac{\Delta \varepsilon}{\hbar}\,,
\label{GP_RCSJ}
\end{eqnarray}
with $U_\mathrm{eff} =| U_R - U_0 p|$ and $s=\mathrm{sign}(U_0p-U_R)$. There is a one-to-one correspondance between these equations and Eq.~\eqref{RCSJ} when one maps $\frac{1}{2U_\mathrm{eff}} \rightarrow \frac{C}{2e^2}$, $\frac{\gamma}{2U_\mathrm{eff}} \rightarrow \frac{1}{2e^2R}$, $-\frac{\gamma}{2U_\mathrm{eff}} \Delta \varepsilon\rightarrow -\frac{\Delta \varepsilon}{2e^2R} \rightarrow I/e$, $-(\frac{\Delta\varepsilon}{2}+z (U_0p-U_R))\rightarrow eV$ and $s\frac{2J}{\hbar} \rightarrow I_c/e$.
When the  energy detuning $\Delta \varepsilon$ is tuned closed to resonance with the confined phonon frequencies, self-induced coherent mechanical oscillations are induced~\cite{Chafatinos2020} and these generate a modulation similar to what microwaves do in a RCSJ. This similarity is demonstrated by comparing panels (b) and (d) of Fig.~\ref{Fig1}. Panel (d) presents numerical calculations using the full gGPEs for the case of modulating either $\Delta \varepsilon(t)$ or $J(t)$, that is,  $\Delta \varepsilon (t) = \Delta \varepsilon_0 + \Delta \varepsilon^{}_\mathrm{M} \cos{(\omega^{}_\mathrm{M} t)}$ or  $J(t) = J_0 +  J^{}_\mathrm{M} \cos{(\omega^{}_\mathrm{M} t)}$.
The phenomena in both cases are qualitatively similar, exhibiting Shapiro steps in  $\langle \dot{\phi} \rangle$ vs. $I_0$ ($I_0/e\equiv\frac{\gamma}{2U_\mathrm{eff}} \Delta \varepsilon$) that lock to harmonics of the driving frequency. As an additional insight, the right plot of Fig.~\ref{Fig1}(d) depicts how the time-averaged population imbalance $\langle z \rangle$ changes as the steps form. This reveals jumps (`Shapiro spikes') whenever $\langle \dot{\phi} \rangle = n \omega_\mathrm{M}$ with $n\in\mathbb{Z}$, evidencing resonant tunneling across the junction under these conditions.

To describe the experiments, numerical calculations of $\langle \dot\phi \rangle$ (a measure of the $\Delta \nu$) as a function of the bare detuning $\Delta \varepsilon$, and of $\langle z \rangle$ as a function of $\langle\dot\phi \rangle$ (normalized to the phonon frequency $\nu_\textrm{M}$), based on the full PJJ gGPEs, are presented in Fig.~\ref{Fig2}d. In this case, we assume that the tunneling rate is modulated as $J(t) = J_0 + J^{(2)}_\textrm{M} \cos{(\omega^{(2)}_\textrm{M} t)}$. Notably, even though $p_1 \neq p_2$ and $\langle z \rangle$ is close to $1$ (contradicting the assumptions made for mapping the coupled gGPEs with the RCSJ model), the system still exhibits a dynamics highly similar to that found in a SSJ under ac modulation, underscoring the robustness of the analogy.

Similar measurements and analysis were conducted for various PJJs with differing dimensions, hence different tunneling rates and unpumped detunings. Two representative experiments are illustrated in Fig.~\ref{Fig3}(a), where $\langle z \rangle$ is plotted against $\Delta \nu$. The presented experiments correspond to coupled square traps with sides measuring $4\,\mu$m and separated either $1\,\mu$m or  $2\,\mu$m. 
Locking and resonant tunneling are clearly observed at different overtones of the confined cavity vibrations, namely at $\sim 20$~GHz, and $\sim 100$~GHz. 


A notable feature in Fig.~\ref{Fig3}(a) is the contrasting shape of the Shapiro spikes corresponding to the self-induced mechanical oscillations at $\sim 20$~GHz and $\sim 100$~GHz. The latter presents a broad asymmetric sawtooth-like behavior, while the former displays a more spiky and narrow one. Figures~\ref{Fig3}(b) and \ref{Fig3}(c) show the calculation using the full gGPEs for the parameters indicated in the caption of the figure. The only difference is that the modulation frequency in (c) is $5$ times larger than in (b), and $\gamma$ has been slightly decreased. We found that larger dissipation further enhances the `spiky' shape as observed in the experiment (for the studied samples, this increase in $\gamma$ can be justified by the lower quality of traps fabricated closer one to the other, at the present technological limit of the technique). Indeed, the two contrasting situations are well reproduced by the model, which then allows to interpret the different dynamics involved in each of them. To this purpose we include in panels (b) and (c) the corresponding trajectories in the phase space ($\phi$ mod $2\pi$, $z$), obtained at the four indicated values of detuning, and spanning a large series of different initial conditions. These phase diagrams are expressed in the rotating frame of the mechanical oscillation. Notably, the difference between the two type of behaviors is that the transition out of the locking (Shapiro step) phase, when decreasing the time averaged dressed detuning $\langle \dot\phi \rangle$, occurs to a single attractor in the spiky case (b) (to a self-trapping situation $\langle z \rangle \sim 1$), while for the sawtooth-like behavior (c) two different attractors can be accessed (either self-trapping or Josephson-like oscillations). These Josephson oscillations are characterized by smaller values of $\langle z \rangle $ and by being strongly locked to the driving mechanical frequency (orbits become a stable point in the rotating wave phase diagram). A dependence on the initial conditions close to locking is also observed in models of synchronization of polariton BECs~\cite{Wouters2008,Ramos2024}. Note also that on crossing the Shapiro resonance the orbits jump a $\sim\pi$ phase. We note that phase slips are ubiquitous in SJJ. They occur when the phase difference across a SJJ suddenly relaxes~\cite{Pop2010,Gumus2023}. These phase slips lead to a sudden change in the current and the generation of a voltage pulse--- here analog to a pulse in $\langle z\rangle$.


 
\textit{Conclusions.---}We have demonstrated Shapiro steps (locking) and spikes (macroscopic tunneling) for coupled exciton polariton condensates under a mechanically induced harmonic time modulation. Notably, these vibrations arise internally through the dynamics of self-induced coherent phonons rather than being externally injected into the cavity. The close relation to the physics of the resistively and capacitively shunted Josephson superconducting junctions was highlighted, suggesting the possible use of the reported phenomena for the development of polariton condensate analogs of quantum circuits~\cite{Barrat2024}. 




\begin{acknowledgments}
We acknowledge partial financial support from the ANPCyT-FONCyT (Argentina) under grants PICT 2019-0371 and PICT 2020-3285. ASK and PVS acknowledge the funding from German DFG (grant 359162958). 
\end{acknowledgments}



\onecolumngrid







\onecolumngrid
\pagebreak

\setcounter{section}{0}
\setcounter{page}{1}
\setcounter{figure}{0}
\renewcommand{\thesection}{Supplementary Material\arabic{section}}
\setcounter{equation}{0}
\renewcommand{\theequation}{S\,\arabic{equation}}

\begin{center}
	\textbf{\Large Supplementary Material: Shapiro steps and spikes in ac-modulated exciton-polariton Josephson junctions}
\end{center}

This supplementary material presents a more detailed derivation of the equivalence between the RCSJ and generalized Gross Pitaevskii models of a junction.\\

\twocolumngrid
\section{Generalized Gross Pitaevskii model of a polariton Josephson junction}


At its core, a Josephson junction comprises two bosonic macro-quantum states, akin to the BCS states found in superconductors, weakly coupled through tunneling. We provide here details of the mapping between the resistively and capacitively shunted junction (RCSJ) model, and the generalized Gross-Pitaevskii (gGPE) equations, which describe the dynamics of polaritons confined in coupled traps.

The weakly coupled modes of the polariton junction are modeled by two gGPEs, which are coupled by a tunneling term~\cite{SM_Wouters2007,SM_CarusottoRMP2013}:
\begin{eqnarray}
	i \hbar \dt{\tilde{\psi}_j} &=& (\varepsilon_j + U_0 |\tilde{\psi}_j|^2 + U_{\textrm{R}} \tilde{n}_j)\tilde{\psi}_j - J\tilde{\psi}_{3-j} \nonumber\\
	&& + \frac{i\hbar\gamma}{2}(\tilde{n}_j - 1)\tilde{\psi}_j,\label{SM_GP}\\
	\dt{\tilde{n}_j} &=& \gamma_R \left[p_j - \left( 1 + |\tilde{\psi}_j|^2  \right) \tilde{n}_j\right].\nonumber
\end{eqnarray}
Here, $\tilde{\psi}_j$ (for $j = 1,2$) represents the amplitude of each trapped mode, while $\tilde{n}_j$ denotes the density of the exciton reservoirs. These variables have been rescaled from their original versions ($\psi_j$ and $n_j$) using the densities $n_0 = \frac{\gamma}{R}$ and $\rho_0 = \frac{\gamma_R}{R}$. Specifically, $\tilde{\psi}_j = \frac{\psi_j}{\sqrt{\rho_0}}$ and $\tilde{n}_j = \frac{n_j}{n_0}$. In this context, $\gamma$ ($\gamma_R$) corresponds to the decay rate of polaritons (excitons) and $R n_j$ represents the filling rate of the condensates from the reservoir.

The term $\varepsilon_j$ corresponds to the bare energy of each mode, $U_0 |\tilde{\psi}_j|^2$ represents the polariton-polariton Coulomb interaction, and $U_{\textrm{R}} \tilde{n}_j$ accounts for the polariton-reservoir-exciton interaction. Additionally, $\frac{J}{\hbar}$ denotes the tunneling rate between the modes.

The equation for $\dt{\tilde{n}_j}$ describes the dynamics of the reservoirs. In this equation, $p_j$ represents the filling rate of the reservoir resulting from non-resonant pumping. To facilitate comparison with the condensation threshold power ($P_{th} = \frac{\gamma \gamma_R}{R}$), $p_j$ has been rescaled as $p_j = \frac{P_j}{P_{th}}$. The term $\left( 1 + |\tilde{\psi}_j|^2  \right) \tilde{n}_j$ accounts for the suppression of reservoir population growth due to exciton decay and decay into the condensates.

As we demonstrate next, through reasonable approximations these equations can be mapped to the RCSJ model. To demonstrate this, consider first the adiabatic approximation for the reservoirs, where $\dt{n_j} = 0$. Secondly, assume that the system is well above the condensation threshold, such that $\left| \tilde{\psi}_j\right|^2 \gg 1$ (relatively large populations with respect to the one at condensation). Then, by using the change of variables $\tilde{\psi}_j = \sqrt{N_j} e^{i\theta_j}$, Eqs. \eqref{SM_GP} reduce to:
\begin{eqnarray}
	\dot{N_j} &=& \gamma \left(p_j - N_j \right) + (-1)^{j+1} \frac{2J}{\hbar} \sqrt{N_1 N_2} \sin{(\phi)} \label{SM_Ntheta}\\
	\dot{\theta_j} &=& -\frac{\varepsilon_j + U_0 N_j + p_j U_R/N_j}{\hbar} + \frac{J}{\hbar} \sqrt{\frac{N_{3-j}}{N_j}} \cos{(\phi)}.\nonumber
\end{eqnarray}
Here, $\phi = \theta_1 - \theta_2$ represents the phase difference between both condensates, analogous to the phase difference in SJJs. Without loss of generality, and to simplify the expressions, let $p_1 = p_2 = p$. 

The four Eqs.\eqref{SM_Ntheta} can be reduced to two. By summing the equations $\dot{N}_1 + \dot{N}_2 = \dot{N}$, one obtains $\dot{N} = \gamma(p_1+p_2 -  N)$. This demonstrates that the total population exponentially approaches a stationary value $N(t=\infty) = 2p$. Thus, this variable can be effectively treated as a constant since the experiments reach a stationary state. Moreover, upon inspecting Eqs. \eqref{SM_Ntheta}, the dynamics between both modes are governed by the relative phase $\phi$ rather than the global phase $\theta_1 + \theta_2$. Consequently, a change of variables to $z = \frac{N_1 - N_2}{N}$ and $\phi$ yields ($\Delta \varepsilon = \varepsilon_1 - \varepsilon_2$): 
\begin{eqnarray}
	\dot{z} &=& - \gamma z + \frac{2J}{\hbar} \sqrt{1-z^2}\sin{(\phi)}\nonumber\\
	\dot{\phi} &=& -\frac{\Delta\varepsilon}{\hbar} - 2\frac{U_0}{\hbar} p z + 2 \frac{U_R}{\hbar} \frac{z}{1-z^2} \\
	&&- \frac{2J}{\hbar} \frac{z}{\sqrt{1-z^2}} \cos{(\phi)}. \nonumber
	\label{SM_zphi}
\end{eqnarray}

Finally, assuming that the condensates are weakly coupled, such that $\left| J \right|\ll \left| \ U_0 p -\ U_R \right|$, one can consider the current between the condensates to be relatively small. In other words, the population imbalance $z \ll 1$. These approximations are standard for superconducting Josephson junctions, from which the following simplified model is derived ($ U_\mathrm{eff} = \left|U_\mathrm{R} - U_0 p \right|$, $s = \mathrm{sign} \left( U_0 p - U_\mathrm{R}\right)$):
\begin{eqnarray}
	\nonumber
	-\frac{\gamma \Delta \varepsilon}{2 U_\mathrm{eff}}  &=& \frac{\hbar}{2U_\mathrm{eff}} \ddot{\phi} + \frac{\hbar \gamma}{2U_\mathrm{eff}} \dot{\phi}    +s \frac{2J}{\hbar} \sin{\phi}\,,\\
	\dot{\phi} &=& 
	-\frac{2(U_0p-U_R)}{\hbar} z - \frac{\Delta \varepsilon}{\hbar}\,,
	\label{GP_RCSJ}
\end{eqnarray}
The first equation above represents precisely the same model as the RCSJ for a SJJ with a driven current. Thus, one can map the parameters of the model to the resulting electrical circuit, where $\frac{1}{2U_\mathrm{eff}} \rightarrow \frac{C}{2e^2}$ plays the role of the capacitance, $\frac{\gamma}{2U_\mathrm{eff}} \rightarrow \frac{1}{2e^2R}$ that of the conductance, $-\frac{\gamma}{2U_{eff}} \Delta \varepsilon \rightarrow I/e$ represents the driven current, and $s\frac{2J}{\hbar} \rightarrow I_c$ can be identified with the critical current in the Josephson junction.

We note on ending, that the approximations used to derive this correspondence between the RCSJ model and the gGPEs are required to obtain the equivalent analytical expressions given above. However, the parallel is much stronger. The calculations presented in the main text for the coupled polariton condensates, which evidence the Shapiro steps and spikes, were obtained using the full gGPEs without any approximation.



\onecolumngrid

\end{document}